\documentclass[journal,letterpaper,twoside,draftcls,onecolumn,12pt]{IEEEtran}%
\usepackage{amsmath}
\usepackage{amssymb}
\usepackage{amsfonts}
\usepackage{cite}
\usepackage{graphicx}%
\providecommand{\U}[1]{\protect\rule{.1in}{.1in}}
\newtheorem{theorem}{Theorem}

\newtheorem{conjecture}[theorem]{Conjecture}

\newtheorem{lemma}[theorem]{Lemma}

\newtheorem{proposition}[theorem]{Proposition}

\allowdisplaybreaks[1]
\begin{document}

\title{Asymptotic Outage Probability Analysis for General Fixed-Gain
Amplify-and-Forward Multihop Relay Systems}
\author{Justin~P.~Coon,~\IEEEmembership{Senior Member,~IEEE,}
Yue~Wang,~\IEEEmembership{Member,~IEEE,}
and~Gillian~Huang,~\IEEEmembership{Student
Member,~IEEE}\thanks{The authors are with Toshiba Research Europe Ltd.,
Telecommunications Research Laboratory, 32 Queen Square, Bristol, BS1 4ND;
tel: +44 (0)117 906 0700, fax: +44 (0)117 906 0701, email: \{justin,yue.wang,gillian.huang\}@toshiba-trel.com.}}
\maketitle

\vspace{-2.0cm}

\begin{abstract}
In this paper, we present an analysis of the outage probability for fixed-gain
amplify-and-forward (AF) multihop relay links operating in the high SNR
regime. Our analysis exploits properties of Mellin transforms to derive an
asymptotic approximation that is accurate even when the per-hop channel gains
adhere to completely different fading models. The main result contained in the
paper is a general expression for the outage probability, which is a
functional of the Mellin transforms of the per-hop channel gains. Furthermore,
we explicitly calculate the asymptotic outage probability for four different
systems, whereby in each system the per-hop channels adhere to either a
Nakagami-$m$, Weibull, Rician, or Hoyt fading profile, but where the
distributional parameters may differ from hop to hop. This analysis leads to
our second main result, which is a semi-general closed-form formula for the
outage probability of general fixed-gain AF multihop systems. We exploit this
formula to analyze an example scenario for a four-hop system where the per-hop
channels follow the four aforementioned fading models, \emph{i.e.}, the first
channel is Nakagami-$m$ fading, the second is Weibull fading, and so on.
Finally, we provide simulation results to corroborate our analysis.

\end{abstract}

\begin{IEEEkeywords}
Amplify-and-forward relaying, semi-blind relaying, fading, outage, Mellin transforms.
\end{IEEEkeywords}

\section{Introduction}

Amplify-and-forward (AF) relay systems have received a lot of attention
recently due to their ability to improve coverage and, thus, capacity in a
geographical sense. To date, two main AF protocols have been focused on in the
literature: variable-gain (a.k.a. channel state information (CSI) assisted) AF
relaying and fixed-gain AF relaying (see, \emph{e.g.},
\cite{Karagiannidis2006} and the references therein). While the former method
yields good performance when CSI is available at the relay nodes, the latter
technique is more suitable in simple systems where such information is not
available, although the performance of the system often suffers. In
particular, fixed-gain AF relaying may be a good choice in low-complexity
systems, such as emerging energy and utility management applications
(\emph{e.g.}, \textquotedblleft smart grid\textquotedblright\ and water
metering communication networks) as well as industrial wireless sensor
networks \cite{Weber2009}. Consequently, we focus on the fixed-gain protocol
in this paper.

As with many wireless communication systems operating in fading environments,
the end-to-end outage probability is an important metric that can be used to
characterize the performance of a fixed-gain AF relaying
system\footnote{Symbol error probability is also an important metric that has
been investigated for fixed-gain AF relaying (see, e.g.,
\cite{Karagiannidis2006a,Farhadi2008,Farhadi2010} and the references
therein).}. Several results on this topic have been published. In
\cite{Karagiannidis2006}, the authors derived a bound on the end-to-end SNR of
a fixed-gain AF link, which was used to study the outage probability when each
hop fades according to a Nakagami-$m$ distribution. This analysis was adapted
and extended in \cite{Karagiannidis2006a} for cases where the per-hop fading
distributions are Nakagami-$n$ (Rice) and Nakagami-$q$ (Hoyt), and the outage
probability was studied using Pad\'{e} approximants. In \cite{Farhadi2008},
the authors derived tight closed-form bounds on the outage probability at
asymptotically high SNR for the case where the underlying channel power
probability density function (PDF) is nonzero at the origin, a condition that
is valid for Rayleigh, Rician, and Hoyt fading, but excludes Nakagami-$m$ (for
$m>1$) and Weibull fading. Other more recent performance analysis results for
multihop relay networks focus on variable-gain AF relaying, particularly in
Nakagami-$m$ fading channels \cite{Amarasuriya2011,Lateef2011}.

There are two main drawbacks to results currently available on fixed-gain AF
relaying, which we aim to improve upon in this paper:

\begin{enumerate}
\item Many of the results found in the literature are given as lower bounds on
the outage probability (see, \emph{e.g.},
\cite{Karagiannidis2006,Karagiannidis2006a,Farhadi2008}). To date, exact
asymptotic results (not bounds) for general multihop systems have not been
reported. Indeed, exact results for dual-hop systems operating in a class of
Nakagami-$m$ channels have only recently been published \cite{DiRenzo2009,Xia2011}.


\item Most analysis of multihop systems to date considers homogeneous fading,
\emph{i.e.}, each hop fades according to the same distribution. Results for
inhomogeneous systems appear to be limited to dual-hop links (see,
\emph{e.g.}, \cite{Suraweera2009a,Suraweera2009}), although this scenario is
likely to be encountered frequently in practical multihop systems.
\end{enumerate}

In this paper, we present a general framework for analyzing the outage
probability of fixed-gain AF multihop relay systems operating in the high SNR
regime. In contrast to outage performance analysis that currently exists in
the literature, our approach does not rely on bounding the outage probability,
but rather exploits properties of Mellin transforms to derive an asymptotic
approximation -- which is in the form of a functional of the Mellin transforms
of the per-hop channel gain PDFs -- that is accurate even when the per-hop
channel gains adhere to completely different fading models. Furthermore, the
nature of the functional allows us to apply the residue theorem from complex
analysis to derive asymptotic approximations for the outage probability
for specific fading models -- including Nakagami-$m$, Weibull, Rician, and
Hoyt fading -- in terms of elementary functions, which can be calculated 
easily in practice. Our contribution culminates
in the introduction of a semi-general asymptotic formula for the outage
probability of fixed-gain AF multihop systems.

The rest of the paper is organized as follows. In
Section II, we define the AF system model. In Section III, the general
framework for the asymptotic outage probability is detailed.  This analysis 
draws heavily on properties of Mellin transforms (outlined in Appendix A for convenience).
In Section IV, the analytical framework is applied to scenarios whereby the per-hop channels adhere to
either a Nakagami-$m$, Weibull, Rician, or Hoyt fading profile, or a
combination thereof. This section concludes with the introduction of a
semi-general asymptotic formula for the outage probability and a brief
discussion on the convergence of the asymptotics. Simulation results that
corroborate our analysis are presented in Section V, and conclusions are
drawn in Section VI.

\section{System Model}

\begin{figure}
[ptb]
\begin{center}
\includegraphics[
height=0.6192in,
width=5.5486in
]%
{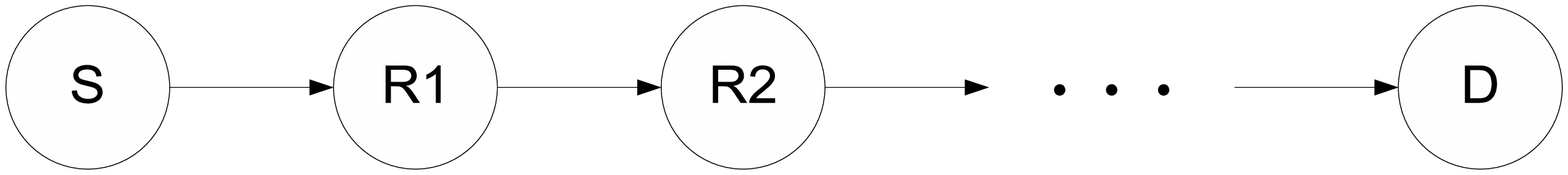}%
\caption{Multihop system diagram with one source transmitting to a destination
via $N-1$ relay nodes; no direct source-destination link exists.}%
\label{fig:system_model}%
\end{center}
\end{figure}

Consider a multihop network with a source node, a destination node, and $N-1$
relay nodes in between where $N\geq2$ (see Fig. \ref{fig:system_model}).
Communication can only be achieved in a half-duplex manner between adjacent
nodes. A data symbol $d$ is conveyed to the first relay node. For simplicity
and without loss of generality, we let $E\left[  \left\vert d\right\vert
^{2}\right]  =1$. This symbol is affected by flat fading in the transmission
medium and additive Gaussian noise at the receiver (\emph{i.e.}, the relay node).

The received signal is then amplified by a fixed gain $A_{1}^{2}$, then
conveyed to the next relay node and so on until the destination is reached.
Denote the channel coefficient modelling the channel between the $\left(
n-1\right)  $th relay (or the source) and the $n$th relay (or the destination)
by $h_{n}$. Also, denote the additive noise at the $n$th relay node (or the
destination) by $v_{n}$, which is zero-mean complex Gaussian distributed with
variance $N_{0,n}/2$ per dimension. Now, we can write the following
input-output system equation \cite{Hasna2003}%
\begin{equation}
r=\left(  \prod_{n=1}^{N}A_{n-1}h_{n}\right)  d+\sum_{n=1}^{N}\left(
\prod_{j=n+1}^{N}A_{j-1}h_{j}\right)  v_{n} \label{eq:sys_model}%
\end{equation}
where $A_{0}=1$. Various amplification factors have been proposed in the
literature, including \cite{Nabar2004}
\begin{equation}
A_{n}=\frac{1}{\sqrt{E\left[  \left\vert h_{n}\right\vert ^{2}\right]
+N_{0,n}}},\quad n=1,\ldots,N-1. \label{eq:An}%
\end{equation}
In order to maintain generality, however, we do not explicitly define $A_{n}$
in what follows.

Since we will eventually be interested in the asymptotics of the outage
probability, it is beneficial to define a reference parameter $\bar{\gamma}$
that gives a notion of average SNR across the $N$-hop link. In this case, we
let $N_{0,n}={\rho_{n}}/{\bar{\gamma}}$ for $n=1,\ldots,N$ where $\left\{
\rho_{n}\right\}  $ are strictly positive (and finite) scaling factors and
$\rho_{1}=1$ for convenience. In order to illustrate the physical definition
of $\bar{\gamma}$, we note that when (\ref{eq:An}) is adopted for the
amplification factors, the average SNR for the $n$th hop is given by
$\bar{\gamma}_{n}=\left(  {E\left[  \left\vert h_{n}\right\vert ^{2}\right]
}/{\rho_{n}}\right)  \bar{\gamma}$. We can now write the following expression
for the instantaneous end-to-end SNR:%
\begin{equation}
\mathsf{SNR}=\frac{\prod_{n=1}^{N}A_{n-1}^{2}\left\vert h_{n}\right\vert ^{2}%
}{\sum_{n=1}^{N}\rho_{n}\prod_{j=n+1}^{N}A_{j-1}^{2}\left\vert h_{j}%
\right\vert ^{2}}\bar{\gamma}. \label{eq:snr}%
\end{equation}

\section{Outage Probability}

For the ease of exposition, we define the random variable $X_{n}=A_{n-1}%
^{2}\left\vert h_{n}\right\vert ^{2}$ in the following analysis. It follows
from (\ref{eq:snr}) that the outage probability of the fixed-gain AF multihop
link can be written as%
\begin{align}
P_{o}  &  =P\left(  \mathsf{SNR}<\gamma_{th}\right) \nonumber\\
&  =P\left(  \prod_{n=1}^{N}X_{n}-\sum_{n=1}^{N-1}\sigma_{n}\prod_{j=n+1}%
^{N}X_{j}<\sigma_{N}\right) \nonumber\\
&  =P\bigg(\Big(\cdots\big(\left(  X_{1}-\sigma_{1}\right)  X_{2}-\sigma
_{2}\big)\cdots\Big)X_{N}<\sigma_{N}\bigg) \label{eq:Pout1}%
\end{align}
where we have defined $\sigma_{n}\left(  \bar{\gamma}\right)  ={\rho_{n}%
\gamma_{th}}/{\bar{\gamma}}$ for brevity. We further define the random
variables%
\begin{equation}
Z_{n}=W_{n}X_{n+1},\quad n=0,\ldots,N-1 \label{eq:Zn}%
\end{equation}
and%
\begin{equation}
W_{n}=W_{n-1}X_{n}-\sigma_{n}>0,\quad n=1,\ldots,N-1
\end{equation}
with $W_{0}\triangleq1$. Note that $W_{n}$ is a conditional random variable in
that it relates to the translation of $Z_{n-1}$ where it is given that
$Z_{n-1}>\sigma_{n}$. Also, it is clear that $W_{n}$ and $X_{n+1}$ are
statistically independent. Now we can apply this same conditioning on
(\ref{eq:Pout1}) recursively to obtain%
\begin{align}
P_{o}  &  =P\left(  Z_{0}\leq\sigma_{1}\right)  +P\left(  Z_{0}>\sigma
_{1}\right)  P\Big(\big(\cdots\left(  W_{1}X_{2}-\sigma_{2}\right)
\cdots\big)X_{N}<\sigma_{N}\Big)\nonumber\\
&  =P\left(  Z_{0}\leq\sigma_{1}\right)  +P\left(  Z_{0}>\sigma_{1}\right)
\Big(P\left(  Z_{1}\leq\sigma_{2}\right)  +P\left(  Z_{1}>\sigma_{2}\right)
\cdots\nonumber\\
&  \qquad\qquad\times\big(P\left(  Z_{N-2}\leq\sigma_{N-1}\right)  +P\left(
Z_{N-2}>\sigma_{N-1}\right)  P\left(  Z_{N-1}<\sigma_{N}\right)
\big)\cdots\Big)\nonumber\\
&  =1-\prod_{n=1}^{N}\mathcal{F}_{Z_{n-1}}\left(  \sigma_{n}\right)  .
\label{eq:Pout2}%
\end{align}
This is a satisfyingly simple exact expression for the outage probability,
although the calculation of the CCDFs of $\left\{  Z_{n}\right\}  $ remain. It
turns out we can construct an elegant lemma using Mellin transforms (see Appendix A) 
that aids this calculation.

\begin{lemma}
\label{lem:1}The Mellin transform of $f_{Z_{n}}$ can be approximated by%
\begin{multline}
M\left[  f_{Z_{n}};s\right]  \approx\frac{1}{\prod_{j=1}^{n}\mathcal{F}%
_{Z_{j-1}}\left(  \sigma_{j}\right)  }\sum_{\ell_{1}=0}^{L_{1}-1}\cdots
\sum_{\ell_{n}=0}^{L_{n}-1}\left(  -1\right)  ^{\sum_{j=1}^{n}\ell_{j}}\\
\times\frac{\Gamma\left(  s\right)  \prod_{j=1}^{n}\sigma_{j}^{\ell_{j}}%
}{\Gamma\left(  s-\sum_{j}\ell_{j}\right)  \prod_{j=1}^{n}\ell_{j}!}%
\prod_{j=1}^{n+1}M\left[  f_{X_{j}};s-%
{\textstyle\sum\nolimits_{k=j}^{n}}
\ell_{k}\right]
\end{multline}
where $L_{1},\ldots,L_{n}\geq1$ are integers that define the order of the approximation.
\end{lemma}

\begin{IEEEproof}
First, we note that $M\left[  f_{Z_{0}};s\right]  =M\left[  f_{X_{1}%
};s\right]  $. Now, from (\ref{eq:Zn}) and property (\ref{eq:property_4}), we
have that $M\left[  f_{Z_{n+1}};s\right]  =M\left[  f_{W_{n+1}};s\right]
M\left[  f_{X_{n+2}};s\right]  $. It is easy to see from the definition of the
random variable $W_{n}$ that $f_{W_{n+1}}\left(  w\right)  =f_{Z_{n}}\left(
w+\sigma_{n+1}\right)  /\mathcal{F}_{Z_{n}}\left(  \sigma_{n+1}\right)  $ for
$w\geq0$. The proof follows from induction on $n$, where properties
(\ref{eq:property_1}) and (\ref{eq:property_2}) are applied in the inductive step.
\end{IEEEproof}

We are now in the position to state our first main result in the form of the
following proposition.

\begin{proposition}
\label{prop:1}The outage probability of an $N$-hop fixed-gain AF link is asymptotically given
by%
\begin{multline}
P_{o}\sim1-\sum_{\ell_{1}=0}^{L_{1}-1}\cdots\sum_{\ell_{N-1}=0}^{L_{N-1}%
-1}\left(  -1\right)  ^{\lambda_{N-1}}\prod_{n=1}^{N-1}\frac{1}{\ell_{n}%
!}\left(  \frac{\rho_{n}}{\rho_{N}}\right)  ^{\ell_{n}}\\
\times\frac{1}{2\pi i}\int_{-\kappa-i\infty}^{-\kappa+i\infty}\left(
\frac{\rho_{N}\gamma_{th}}{\bar{\gamma}}\right)  ^{s}\frac{\Gamma\left(
\lambda_{N-1}-s\right)  }{\Gamma\left(  1-s\right)  }\prod_{n=1}^{N}M\left[
f_{X_{n}};1+\lambda_{n-1}-s\right]  \mathrm{d}s \label{eq:Pout}%
\end{multline}
as $\bar{\gamma}\rightarrow\infty$ with $\kappa>\lambda_{N-1}$, where
$\lambda_{n}=%
{\textstyle\sum\nolimits_{k=1}^{n}}
\ell_{k}$ and $f_{X_{n}}$ denotes the PDF of the channel power for the $n$th hop\footnote{The notation $\sim$ denotes
asymptotic equivalence in the relevant variable and limit, $\bar{\gamma} \rightarrow \infty$ in this case.}.
\end{proposition}

\begin{IEEEproof}
Starting from Lemma \ref{lem:1}, we use (\ref{eq:property_3}) to calculate
$M\left[  \mathcal{F}_{Z_{N-1}};s\right]  $, then take the inverse transform
using (\ref{eq:invM}) and substitute this into (\ref{eq:Pout2}), resulting in
a cancellation of the product $\prod_{n=1}^{N-1}\mathcal{F}_{Z_{n-1}}\left(
\sigma_{n}\right)  $. Eq. (\ref{eq:Pout}) follows from a change of variables
in the contour integral.
\end{IEEEproof}

Although this appears to be a rather complicated expression for $P_{o}$, we
note that it can be computed as long as we know the Mellin transforms of the
individual PDFs $f_{X_{n}}$, which can generally be calculated easily for many
cases of interest. Thus, assuming we can compute these transforms, we can at
least evaluate the outage probability bound numerically to a high degree of
accuracy provided the transforms decay quickly as $\left\vert
\operatorname{Im}\left(  s\right)  \right\vert \rightarrow\infty$.

The Mellin transforms for the main fading distributions of interest are given
in Table \ref{tab:mellin}\footnote{The transforms for the Rician and Hoyt
distributions can be calculated by consulting standard tables of integrals and
transforms (\emph{e.g.}, \cite{Erdelyi1954,Gradshteyn2000}).}. The key thing
to notice from this table that makes the ensuing analysis uniform and
tractable is that each PDF decays exponentially. In the transform domain, this
results in the property that the transform of each fading distribution has
poles at various points along the real axis. We will exploit this property in
the next section to derive simple asymptotic expressions for $P_{o}$ under
various fading conditions.

\section{Analysis of Common Fading Distributions}

We now apply the general analysis detailed above to a number of common fading
distributions. For much of this section, we assume that all hops adhere to the
same class of distribution (\emph{e.g.}, Nakagami-$m$) but where the channels
for individual hops may vary in their distributional parameters (\emph{e.g.},
shape factors). We label this condition \textquotedblleft
homogeneity\textquotedblright\ in this context, and we later relax this
restriction in order to analyze general multihop links as well as to show the
power and versatility of adopting the proposed analytical framework. The
distributions considered here (also listed in Table \ref{tab:mellin}) are
derived from Nakagami-$m$, Weibull, Rician, and Hoyt fading. The analysis can
be extended to other fading distributions using (\ref{eq:Pout}) and the
techniques outlined below. The section concludes with our second major result,
which is in the form of a semi-general closed-form formula for the outage
probability for large $\bar{\gamma}$.

\subsection{Nakagami-$m$ Fading\label{sec:Nakm}}

If the per hop channels adhere to a Nakagami-$m$ fading profile, the random
variable $X_{n}$ has a gamma density function with scale parameter $\theta
_{n}$ and shape parameter $m_{n}$. The corresponding PDF and Mellin transform
are given in Table \ref{tab:mellin}. By substituting this transform into
(\ref{eq:Pout}) and applying the Mellin-Barnes integral definition of the
Meijer $G$-function \cite{Gradshteyn2000} along with the functional relations
\cite[9.31.2]{Gradshteyn2000} and \cite[9.31.5]{Gradshteyn2000}, we can write
the outage probability as%
\begin{equation}
P_{o,Nak}\sim1-\sum_{\ell_{1},\ldots,\ell_{N-1}}\left(  -1\right)
^{\lambda_{N-1}}\xi_{\boldsymbol{\ell}}G_{1,N+1}^{N+1,0}\left(  \frac
{\sigma_{N}}{\prod_{j=1}^{N}\theta_{j}}\left\vert
\begin{array}
[c]{c}%
1\\
\lambda_{N-1},\,m_{1},\,m_{2}+\lambda_{1},\ldots,\,m_{N}+\lambda_{N-1}%
\end{array}
\right.  \right)  \label{eq:Pout_Nakm_full}%
\end{equation}
where%
\[
\xi_{\boldsymbol{\ell}}=\frac{1}{\Gamma\left(  m_{N}\right)  }\prod
_{n=1}^{N-1}\frac{\theta_{n+1}^{\lambda_{n}}\left(  \frac{\rho_{n}}{\rho_{N}%
}\right)  ^{\ell_{n}}}{\ell_{n}!\Gamma\left(  m_{n}\right)  }%
\]
with $\boldsymbol{\ell}=\left(  \ell_{1},\ldots,\ell_{N-1}\right)  $ and
$\lambda_{n}=%
{\textstyle\sum\nolimits_{k=1}^{n}}
\ell_{k}$ (defining $\lambda_{0}=0$). It is understood that the summation in
(\ref{eq:Pout_Nakm_full}) is $\left(  N-1\right)  $-fold, with the sum over
the $n$th index $\ell_{n}$ running from zero to $L_{n}-1$. This expression can
be evaluated easily using mathematical software tools such as Mathematica.

We may wish to consider the leading order expansion of $P_{o,Nak}$ at high
SNR, which would yield expressions for the diversity and coding gains of the
multihop link. To do this, we first note that the Meijer $G$-function used
here is defined as a Mellin-Barnes integral where the integration path goes
from $-i\infty$ to $i\infty$ such that it separates the poles of the integrand
\cite[\S 16.17]{NIST2010}. It is straightfoward to show that this integral
converges in our application and that the integrand, given by%
\begin{equation}
\label{eq:I_Nak}I_{Nak}\left(  s\right)  =z\left(  \bar{\gamma}\right)
^{s}\frac{\Gamma\left(  \lambda_{N-1}-s\right)  \prod_{n=1}^{N}\Gamma\left(
m_{n}+\lambda_{n-1}-s\right)  }{\Gamma\left(  1-s\right)  }%
\end{equation}
where $z\left(  \bar{\gamma}\right)  =\left(  \rho_{N}\gamma_{th}/\prod
_{n}\theta_{n}\right)  \bar{\gamma}^{-1}$, is well-behaved as $s\rightarrow
\infty$ in the right half $s$-plane. Moreover, $I_{Nak}\left(  s\right)  $ has
poles at $s=m_{n}+\lambda_{n-1}+j$ for $n=0,\ldots,N-1$ and $j=0,1,\ldots$,
and at $s=0$ when $\lambda_{N-1}=0$. Thus, we can employ the residue
theorem\footnote{See, \emph{e.g.}, \cite{Stewart1983} for an introduction to
complex analysis and the residue theorem.} with the usual closing arc in the
right half plane to evaluate the $G$-functions in (\ref{eq:Pout_Nakm_full}) at
high SNR, which leads to a more accessible and intuitive asymptotic expression
for the outage probability given by\footnote{The notation $g\left(
\bar{\gamma}\right)  =o\left(  h\left(  \bar{\gamma}\right)  \right)  $
signifies that $\lim_{\bar{\gamma}\rightarrow\infty}g\left(  \bar{\gamma
}\right)  /h\left(  \bar{\gamma}\right)  =0$.}%
\begin{multline}
P_{o,Nak}\sim\frac{\left(  \rho_{N}\gamma_{th}\right)  ^{m}}{\prod_{n}%
\theta_{n}^{m}}\sum_{r=1}^{\mu}\sum_{\left\{  \ell_{1},\ldots,\ell
_{N-1}\right\}  \in\mathcal{L}_{r}}\left(  -1\right)  ^{\lambda_{N-1}}%
\xi_{\boldsymbol{\ell}}\sum_{l=0}^{r-1}\frac{\nu^{\left(  r-1-l\right)
}\left(  m\right)  }{\left(  r-1-l\right)  !l!}\\
\times\sum_{p=0}^{l}\binom{l}{p}\left(  -1\right)  ^{p}\left(  \log\frac
{\rho_{N}\gamma_{th}}{\prod_{n}\theta_{n}}\right)  ^{l-p}\frac{\left(
\log\bar{\gamma}\right)  ^{p}}{\bar{\gamma}^{m}}+o\left(  \bar{\gamma}%
^{-m}\right)  \label{eq:Pout_Nakm_expansion}%
\end{multline}
where $m=\min\left\{  m_{n}\right\}  $, $\mu$ denotes the multiplicity of $m$,
and%
\[
\nu\left(  s\right)  =\left(  s-m\right)  ^{r}\frac{I_{Nak}\left(  s\right)
}{z^{s}}.
\]
The sets $\mathcal{L}_{1},\ldots,\mathcal{L}_{\mu}$ are disjoint sets of
$\left(  N-1\right)  $-tuples $\left\{  \ell_{1},\ldots,\ell_{N-1}\right\}  $
defined such that $I_{Nak}\left(  s\right)  $ has an $r$th order pole at $s=m$
when $\left\{  \ell_{1},\ldots,\ell_{N-1}\right\}  \in\mathcal{L}_{r}$. The
details of the calculations that lead to this result (and a rigorous
definition of the sets $\mathcal{L}_{1},\ldots,\mathcal{L}_{\mu}$) are given
in the appendix.

This expression can be evaluated easily for specific examples, but by
retaining only the leading order term (\emph{i.e.}, $r=\mu$), we arrive at the
following general asymptotic equivalence:%
\begin{equation}
P_{o,Nak}\sim\psi_{Nak}\frac{\left(  \log\bar{\gamma}\right)  ^{\mu-1}}%
{\bar{\gamma}^{m}} \label{eq:Nak_leading_order1}%
\end{equation}
where%
\begin{equation}
\psi_{Nak}=\frac{\left(  \rho_{N}\gamma_{th}\right)  ^{m}}{\prod_{n}\theta
_{n}^{m}}\sum_{\left\{  \ell_{1},\ldots,\ell_{N-1}\right\}  \in\mathcal{L}%
_{\mu}}\left(  -1\right)  ^{\lambda_{N-1}+\mu-1}\frac{\xi_{\boldsymbol{\ell}%
}\nu\left(  m\right)  }{\left(  \mu-1\right)  !} \label{eq:Nak_leading_order2}%
\end{equation}
is the coding gain of the link\footnote{This definition of the coding gain is
slightly different to the one typically used in system analysis. In fact, the
standard definition of the coding gain (see, \emph{e.g.}, \cite{Wang2003a})
cannot be applied here since it is only valid when $P_{out}$ obeys a power law
decay in $\bar{\gamma}$.}. For the case where each hop fades independently of
others and all hops experience nonidentically shaped fading (\emph{i.e.},
$m_{1}\neq\cdots\neq m_{N}$), $\mu=1$ and we have%
\begin{equation}
P_{o,Nak}\sim\frac{\left(  \rho_{N}\gamma_{th}\right)  ^{m}}{\prod_{n}%
\theta_{n}^{m}}\sum_{\left\{  \ell_{1},\ldots,\ell_{N-1}\right\}
\in\mathcal{L}_{\mu}}\left(  -1\right)  ^{\lambda_{N-1}}\xi_{\boldsymbol{\ell
}}\nu\left(  m\right)  \bar{\gamma}^{-m}.
\end{equation}
In the other extreme where all hops experience identically shaped fading
(\emph{i.e.}, $m=m_{1}=\cdots=m_{N}$), $\mu=N$ and we can write%
\begin{equation}\label{eq:id_sh_fading}
P_{o,Nak}\sim\frac{\left(  \rho_{N}\gamma_{th}\right)  ^{m}}{\left(
N-1\right)  !m\Gamma\left(  m\right)  ^{N}\prod_{n}\theta_{n}^{m}}%
\frac{\left(  \log\bar{\gamma}\right)  ^{N-1}}{\bar{\gamma}^{m}}.
\end{equation}

The expressions given above demonstrate the well-known fact that $m$th order
diversity is achieved in these cases. However, the analysis is useful in
illustrating the \emph{rate} (with respect to SNR growth) at which $m$th order diversity is attained.
By taking the (usual) definition of diversity to be
\begin{equation}
d = \lim_{\bar{\gamma} \rightarrow \infty} \frac{\log P_{o,Nak}(\bar{\gamma})}{-\log \bar{\gamma}}	
\end{equation}
we see from (\ref{eq:Nak_leading_order1}) that, for finite but large values of $\bar{\gamma}$,
\begin{equation}
	d\left(\bar{\gamma}\right) = m - (\mu -1)\frac{\log\log\bar{\gamma}}{\log\bar{\gamma}} + O\left(\frac{1}{\log\bar{\gamma}}\right).
\end{equation}

Although this result points to slow convergence in terms of diversity (due to the $\log\log\bar{\gamma}/\log\bar{\gamma}$ term), it does not illustrate
the full picture since the coding gain must be taken into account.  Indeed, through our analysis, we
have presented an accurate expression for the coding gain of a fixed-gain
AF multihop link, which to the best of the authors' knowledge has not yet been
reported in the literature. The coding gain is, in general, a complicated expression,
although it is straightforward to compute.
However, we can draw some conclusions about the behavior of certain systems, such as those
that experience identically shaped fading.  For example, (\ref{eq:id_sh_fading}) points to the importance of 
having a well-designed destination receiver with a high-performance low-noise amplifier (LNA) 
when operating in the high SNR regime ($\rho_N/\bar{\gamma} = N_{0,N}$
must be as low as possible).

Finally, it should be noted that the leading order
results given here are not very accurate for low to mid-range $\bar{\gamma}$
when $\mu>1$. This results from the fact that $\log\bar{\gamma}$ increases
very slowly, which effectively means that all $\left(  \log\bar{\gamma
}\right)  ^{p}\bar{\gamma}^{-m}$ terms are of roughly the same order for
finite $\bar{\gamma}$ and, thus, should be included in the approximation. In
such a case, it is best to use the general expansion given by
(\ref{eq:Pout_Nakm_expansion}).

\subsection{Weibull Fading\label{sec:Wei}}

If the channels adhere to a Weibull fading profile, the random variable
$X_{n}$ has a Weibull density function. The corresponding PDF and Mellin
transform are given in Table \ref{tab:mellin}. By substituting this transform
into (\ref{eq:Pout}) and applying the definition of the Fox $H$-function
\cite{Braaksma1962} along with the functional relations \cite[(2.8)]%
{Mainardi2005} and \cite[(2.11)]{Mainardi2005}, we can write the outage
probability as%
\begin{equation}
P_{o,Wei}\sim1-\sum_{\ell_{1},\ldots,\ell_{N-1}}\left(  -1\right)
^{\lambda_{N-1}}\varphi_{\boldsymbol{\ell}}H_{1,N+1}^{N+1,0}\left(
\frac{\sigma_{N}}{\prod_{j=1}^{N}\theta_{j}}\left\vert
\begin{array}
[c]{c}%
\left(  1,1\right) \\
\left(  \lambda_{N-1},1\right)  ,\left(  1+\frac{\lambda_{0}}{m_{1}},\frac
{1}{m_{1}}\right)  ,\ldots,\left(  1+\frac{\lambda_{N-1}}{m_{N}},\frac
{1}{m_{N}}\right)
\end{array}
\right.  \right)  \label{eq:Pout_Wei_full}%
\end{equation}
where%
\[
\varphi_{\boldsymbol{\ell}}=\prod_{n=1}^{N-1}\frac{\theta_{n+1}^{\lambda_{n}}%
}{\ell_{n}!}\left(  \frac{\rho_{n}}{\rho_{N}}\right)  ^{\ell_{n}}.
\]

As with the Nakagami-$m$ case, we can employ the residue theorem to obtain a
simple asymptotic expression for $P_{o,Wei}$ using elementary functions.
Omitting the details (the methodology is the same as was described for
Nakagami-$m$ fading in the appendix), it is possible to derive the following
asymptotic expression:%
\begin{multline}
P_{o,Wei}\sim\frac{\left(  \rho_{N}\gamma_{th}\right)  ^{m}}{\prod_{n}%
\theta_{n}^{m}}\sum_{r=1}^{\mu}\sum_{\left\{  \ell_{1},\ldots,\ell
_{N-1}\right\}  \in\mathcal{L}_{r}}\left(  -1\right)  ^{\lambda_{N-1}}%
\varphi_{\boldsymbol{\ell}}\sum_{l=0}^{r-1}\frac{\omega^{\left(  r-1-l\right)
}\left(  m\right)  }{\left(  r-1-l\right)  !l!}\\
\times\sum_{p=0}^{l}\binom{l}{p}\left(  -1\right)  ^{p}\left(  \log\frac
{\rho_{N}\gamma_{th}}{\prod_{n}\theta_{n}}\right)  ^{l-p}\frac{\left(
\log\bar{\gamma}\right)  ^{p}}{\bar{\gamma}^{m}}+o\left(  \bar{\gamma}%
^{-m}\right)  \label{eq:Pout_Wei_expansion}%
\end{multline}
where again $m=\min\left\{  m_{n}\right\}  $, $\mu$ denotes the multiplicity
of $m$, and%
\[
\omega\left(  s\right)  =\left(  s-m\right)  ^{r}\frac{\Gamma\left(
\lambda_{N-1}-s\right)  \prod_{n=1}^{N}\Gamma\left(  1+\frac{\lambda_{n-1}%
-s}{m_{n}}\right)  }{\Gamma\left(  1-s\right)  }.
\]
To leading order in $\bar\gamma$, we have the asymptotic equivalence%
\begin{equation}
P_{o,Wei}\sim\psi_{Wei}\frac{\left(  \log\bar{\gamma}\right)  ^{\mu-1}}%
{\bar{\gamma}^{m}}%
\end{equation}
where%
\begin{equation}
\psi_{Wei}=\frac{\left(  \rho_{N}\gamma_{th}\right)  ^{m}}{\prod_{n}\theta
_{n}^{m}}\sum_{\left\{  \ell_{1},\ldots,\ell_{N-1}\right\}  \in\mathcal{L}%
_{\mu}}\left(  -1\right)  ^{\lambda_{N-1}+\mu-1}\frac{\varphi
_{\boldsymbol{\ell}}\omega\left(  m\right)  }{\left(  \mu-1\right)  !}%
\end{equation}
is the coding gain.  Note that this analysis points to similar diversity and coding
gain behavior as was discussed for Nakagami-$m$ fading channels.

\subsection{Rician and Hoyt Fading\label{sec:Hoyt}}

No closed form expression of $P_{o}$ exists for the cases where all hops
follow a Rician (or Hoyt) fading model. For the Rician case, we can employ the
transform given in Table \ref{tab:mellin} along with Proposition \ref{prop:1}
to obtain an expression for $P_{o,Rice}$ in terms of a contour integral, which
can be evaluated numerically using, for example, Mathematica. The residue
theorem can be applied to this integral to derive an asymptotic expression in
the form of%
\begin{multline}
P_{o,Rice}\sim\frac{\rho_{N}\gamma_{th}\prod_{n}\left(  K_{n}+1\right)
}{\prod_{n}\theta_{n}}\sum_{r=1}^{N}\sum_{\left\{  \ell_{1},\ldots,\ell
_{N-1}\right\}  \in\mathcal{L}_{r}}\left(  -1\right)  ^{\lambda_{N-1}}%
\zeta_{\mathbf{\ell}}\sum_{l=0}^{r-1}\frac{\varrho^{\left(  r-1-l\right)
}\left(  1\right)  }{\left(  r-1-l\right)  !l!}\\
\times\sum_{p=0}^{l}\binom{l}{p}\left(  -1\right)  ^{p}\left(  \log\frac
{\rho_{N}\gamma_{th}\prod_{n}\left(  K_{n}+1\right)  }{\prod_{n}\theta_{n}%
}\right)  ^{l-p}\frac{\left(  \log\bar{\gamma}\right)  ^{p}}{\bar{\gamma}%
}+o\left(  \bar{\gamma}^{-1}\right)  \label{eq:Pout_Rice_expansion}%
\end{multline}
where%
\[
\zeta_{\mathbf{\ell}}=e^{-K_{N}}\prod_{n=1}^{N-1}\frac{e^{-K_{n}}}{\ell_{n}%
!}\left(  \frac{\rho_{n}}{\rho_{N}}\right)  ^{\ell_{n}}\left(  \frac
{\theta_{n+1}}{K_{n+1}+1}\right)  ^{\lambda_{n}}%
\]
and%
\[
\varrho\left(  s\right)  =\left(  s-1\right)  ^{r}\frac{\Gamma\left(
\lambda_{N-1}-s\right)  }{\Gamma\left(  1-s\right)  }\prod_{n=1}^{N}%
\Gamma\left(  1+\lambda_{n-1}-s\right)  \,_{1}F_{1}\left(  1+\lambda
_{n-1}-s,1;K_{n}\right)  .
\]
To leading order in $\bar\gamma$, we have%
\begin{equation}
P_{o,Rice}\sim\frac{\rho_{N}\gamma_{th}}{\left(  N-1\right)  !}\prod_{n=1}%
^{N}\frac{K_{n}+1}{\theta_{n}e^{K_{n}}}\frac{\left(  \log\bar{\gamma}\right)
^{N-1}}{\bar{\gamma}}.
\end{equation}

Similarly, for the case where all hops follow a Hoyt fading model, we can
follow the same procedure to derive the asymptotic expression%

\begin{multline}
P_{o,Hoyt}\sim\frac{\rho_{N}\gamma_{th}\prod_{n}\left(  1+q_{n}^{2}\right)
^{2}}{\prod_{n}4q_{n}^{2}\theta_{n}}\sum_{r=1}^{N}\sum_{\left\{  \ell
_{1},\ldots,\ell_{N-1}\right\}  \in\mathcal{L}_{r}}\left(  -1\right)
^{\lambda_{N-1}}\chi_{\boldsymbol{\ell}}\sum_{l=0}^{r-1}\frac{\eta^{\left(
r-1-l\right)  }\left(  1\right)  }{\left(  r-1-l\right)  !l!}\\
\times\sum_{p=0}^{l}\binom{l}{p}\left(  -1\right)  ^{p}\left(  \log\frac
{\rho_{N}\gamma_{th}\prod_{n}\left(  1+q_{n}^{2}\right)  ^{2}}{\prod_{n}%
4q_{n}^{2}\theta_{n}}\right)  ^{l-p}\frac{\left(  \log\bar{\gamma}\right)
^{p}}{\bar{\gamma}} \label{eq:Pout_Hoyt_expansion}%
\end{multline}
where%
\[
\chi_{\boldsymbol{\ell}}=\left(  \frac{2q_{N}}{1+q_{N}^{2}}\right)
^{1+2\lambda_{N-1}}\prod_{n=1}^{N-1}\frac{\theta_{n+1}^{\lambda_{n}}}{\ell
_{n}!}\left(  \frac{\rho_{n}}{\rho_{N}}\right)  ^{\ell_{n}}\left(
\frac{2q_{n}}{1+q_{n}^{2}}\right)  ^{1+2\lambda_{n-1}}%
\]
and%
\[
\eta\left(  s\right)  =\left(  s-1\right)  ^{r}\frac{\Gamma\left(
\lambda_{N-1}-s\right)  }{\Gamma\left(  1-s\right)  }\prod_{n=1}^{N}%
\Gamma\left(  1+\lambda_{n-1}-s\right)  \,_{2}F_{1}\left(  \frac
{1+\lambda_{n-1}-s}{2},\frac{2+\lambda_{n-1}-s}{2};1;\left(  \frac{1-q_{n}%
^{2}}{1+q_{n}^{2}}\right)  ^{2}\right)  .
\]
To leading order in $\bar\gamma$, (\ref{eq:Pout_Hoyt_expansion}) reduces to%
\begin{equation}
P_{o,Hoyt}\sim\frac{\rho_{N}\gamma_{th}}{\left(  N-1\right)  !}\prod_{n=1}%
^{N}\frac{1+q_{n}^{2}}{2q_{n}\theta_{n}}\frac{\left(  \log\bar{\gamma}\right)
^{N-1}}{\bar{\gamma}}.
\end{equation}

\subsection{Semi-general Formula\label{sec:semi-general}}

It is clear that similarities exist between the expressions for $P_{o}$ given
for the different fading distributions analyzed above. This observation leads
to our second main result, which is in the form of a semi-general asymptotic
formula for the outage probability of fixed-gain AF multihop links.

\begin{conjecture}
\label{conj:1}Consider an $N$-hop fixed-gain AF link, where the PDF of the
channel power for each hop decays exponentially, and thus has a Mellin
transform that is well-behaved at $\left\vert \operatorname{Im}\left(
s\right)  \right\vert =\infty$. The outage probability of this link is given
by%
\begin{multline}
P_{o}\sim A^{m}\sum_{r=1}^{\mu}\sum_{\left\{  \ell_{1},\ldots,\ell
_{N-1}\right\}  \in\mathcal{L}_{r}}\left(  -1\right)  ^{\lambda_{N-1}%
}B_{\boldsymbol{\ell}}\sum_{l=0}^{r-1}\frac{C_{r}^{\left(  r-1-l\right)
}\left(  m\right)  }{\left(  r-1-l\right)  !l!}\\
\times\sum_{p=0}^{l}\binom{l}{p}\left(  -1\right)  ^{p}\left(  \log A\right)
^{l-p}\frac{\left(  \log\bar{\gamma}\right)  ^{p}}{\bar{\gamma}^{m}}+o\left(
\bar{\gamma}^{-m}\right)  \label{eq:conj1}%
\end{multline}
as $\bar{\gamma}\rightarrow\infty$. In (\ref{eq:conj1}), the constants $A$ and
$B_{\boldsymbol{\ell}}$ are dependent upon the distributional parameters for
the $N$ channels in the link with $B_{\boldsymbol{\ell}}$ being dependent upon
the indices in the vector $\boldsymbol{\ell}=\left(  \ell_{1},\ldots
,\ell_{N-1}\right)  $ as well; $m$ defines the minimum shape parameter of the
$N$ channel power distributions with $\mu$ being the multiplicity of $m$
(\emph{e.g.}, $m=1$ and $\mu=N$ if all hops are Rayleigh, Rician, or Hoyt);
and $C_{r}\left(  s\right)  =\left(  s-m\right)  ^{r}I\left(  s\right)
\left(  \bar{\gamma}/A\right)  ^{s}$ with $I\left(  s\right)  $ being a simple
form of the integrand in (\ref{eq:Pout}) where all terms independent of $s$
have been removed.
\end{conjecture}

This result is stated as a conjecture since we make no attempt to rigorously
define the properties that the channel power PDFs must have to make it a
theorem. Indeed, only a few channel models are typically employed in practice.
These models include those discussed in Sections \ref{sec:Nakm} through
\ref{sec:Hoyt}. We can easily apply the formula to those scenarios.

Additionally, this formula can be used to characterize the outage probability
of inhomogeneous links. As a toy example, which is perhaps not often
encountered in practice but serves as a useful illustration, consider a
four-hop link where the per-hop channels are consecutively modelled as
experiencing Nakagami-$m$, Weibull, Rician, and Hoyt fading. In this case, it
is straightforward to show that the semi-general formula given by
(\ref{eq:conj1}) holds with the following definitions:%
\begin{align*}
A  &  =\frac{\rho_{N}\gamma_{th}\left(  K_{3}+1\right)  \left(  1+q_{4}%
^{2}\right)  ^{2}}{4q_{4}^{2}\prod_{n}\theta_{n}}\\
B_{\mathbf{\ell}}  &  =\frac{e^{-K_{3}}}{\Gamma\left(  m_{1}\right)  \left(
K_{3}+1\right)  ^{\lambda_{2}}}\left(  \frac{2q_{4}}{1+q_{4}^{2}}\right)
^{1+2\lambda_{3}}\prod_{n=1}^{N-1}\frac{\theta_{n+1}^{\lambda_{n}}}{\ell_{n}%
!}\left(  \frac{\rho_{n}}{\rho_{N}}\right)  ^{\ell_{n}}\\
C_{r}\left(  s\right)   &  =\left(  s-m\right)  ^{r}\frac{\Gamma\left(
\lambda_{N-1}-s\right)  }{\Gamma\left(  1-s\right)  }\Gamma\left(
m_{1}-s\right)  \Gamma\left(  1+\frac{\lambda_{1}-s}{m_{2}}\right)
\Gamma\left(  1+\lambda_{2}-s\right)  \Gamma\left(  1+\lambda_{3}-s\right) \\
&  \quad\times\,_{1}F_{1}\left(  1+\lambda_{2}-s,1;K_{3}\right)  \,_{2}%
F_{1}\left(  \frac{1+\lambda_{3}-s}{2},\frac{2+\lambda_{3}-s}{2};1;\left(
\frac{1-q_{4}^{2}}{1+q_{4}^{2}}\right)  ^{2}\right) \\
m  &  =\min\left\{  m_{1},m_{2},1\right\}  .
\end{align*}
This example is discussed further in Section \ref{sec:results}.

\subsection{Convergence of Asymptotics\label{sec:convergence}}

It is natural to determine the conditions under which the leading order
expressions for $P_{o}$ given in the preceding section serve as good
approximations to the actual outage probability. First, we study the question:
how large does $\bar{\gamma}$ have to be before \textquotedblleft
asymptotic\textquotedblright\ becomes \textquotedblleft
approximate\textquotedblright? For nonidentical Nakagami-$m$ and Weibull
fading scenarios, convergence occurs quickly since the leading order
expression monotonically decreases with increasing $\bar{\gamma}$. For
quasi-identical fading, as well as Rician and Hoyt fading channels, we note
that the asymptotic bound for $P_{o}$ given above has the form $P_{o}\sim
b\left(  \log\bar{\gamma}\right)  ^{\mu-1}\bar{\gamma}^{-m}$ where $b$ is
independent of $\bar{\gamma}$, $\mu\geq2$, and $m\geq1$. For $\bar{\gamma}>1$,
it can be shown that this expression has a maximum at $\bar{\gamma}=e^{\left(
\mu-1\right)  /m}$, and is monotonically decreasing for $\bar{\gamma
}>e^{\left(  \mu-1\right)  /m}$. Thus, it is necessary that $\bar{\gamma}\gg
e^{\left(  \mu-1\right)  /m}$ for the asymptotic expressions given above to be
reasonable approximations for the outage probability. For hardened channels
(\emph{i.e.}, when $m\gg1$ for Nakagami-$m$ and Weibull fading) or systems
with only a few hops, this condition is satisfied easily in practice.

\section{Simulation Results and Discussion\label{sec:results}}

In this section, we present numerical results obtained for homogeneous and
inhomogeneous multihop links. For the homogeneous case, we simulated a number
of scenarios in order to validate our analytical results. All four fading
distributions mentioned above were considered, and different system parameters
were chosen to illustrate the accuracy of our analytical framework as well as
some interesting behavior of fixed-gain multihop links. For the inhomogeneous
case, the four-hop example discussed in Section \ref{sec:semi-general} was
studied. For all calculations, we assume only a first order asymptotic
correction, \emph{i.e.}, $L_{n}=2$ for $n=1,\ldots,N-1$.

\begin{figure}
[ptb]
\begin{center}
\includegraphics[
height=4.0179in,
width=5.1906in
]%
{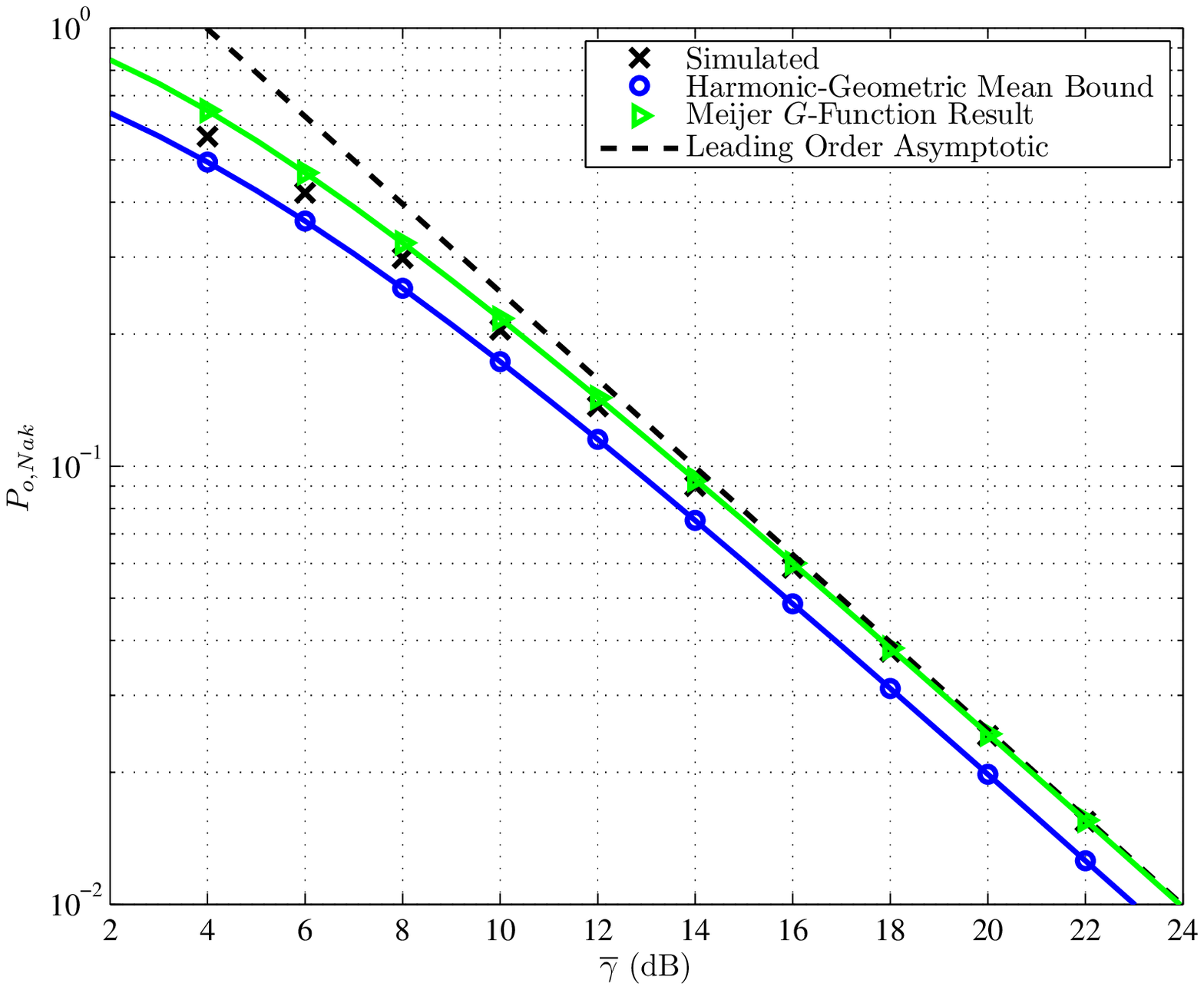}%
\caption{$P_{o,Nak}$ vs. $\bar{\gamma}$ for a fixed-gain AF multihop systems
with Nakagami-$m$ fading channels ($N=3$; $m_{n}=n$, $K_{n}=2$ and $\theta
_{n}=\rho_{n}=1$ for all $n$).}%
\label{fig:Nak_fig1}%
\end{center}
\end{figure}

\begin{figure}
[ptb]
\begin{center}
\includegraphics[
height=4.0179in,
width=5.1906in
]%
{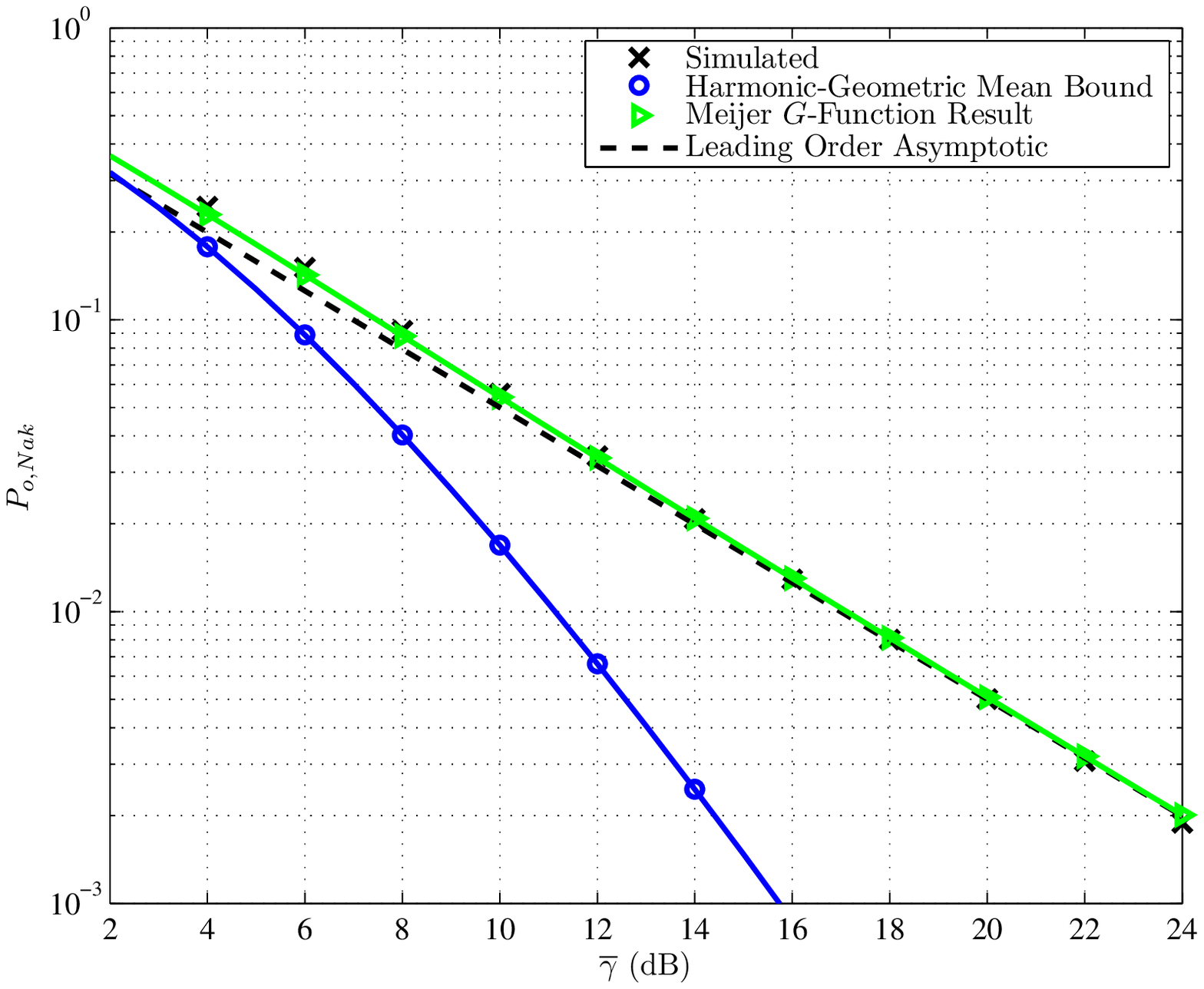}%
\caption{$P_{o,Nak}$ vs. $\bar{\gamma}$ for a fixed-gain AF multihop systems
with Nakagami-$m$ fading channels ($N=3$; $m_{n}=N-n+1$, $K_{n}=2$, and
$\theta_{n}=\rho_{n}=1$ for all $n$).}%
\label{fig:Nak_fig2}%
\end{center}
\end{figure}

\begin{figure}
[ptb]
\begin{center}
\includegraphics[
height=4.0179in,
width=5.1906in
]%
{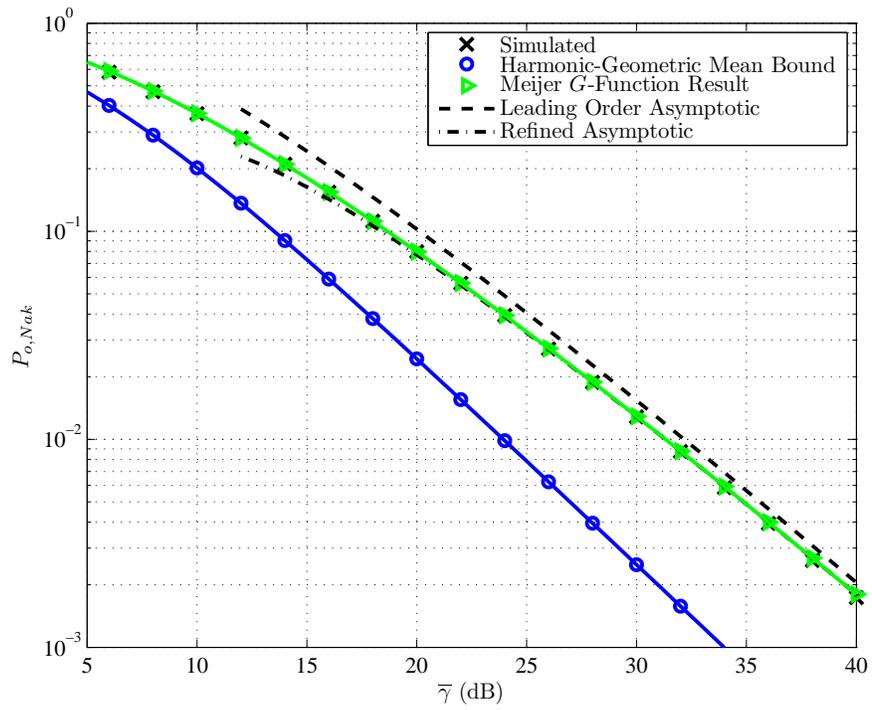}%
\caption{$P_{o,Nak}$ vs. $\bar{\gamma}$ for a fixed-gain AF multihop systems
with Nakagami-$m$ fading channels ($N=3$, $m_{1}=m_{3}=1$, $m_{2}=2$,
$\rho_{1}=1$, $\rho_{2}=1/3$, $\rho_{3}=5/3$; and $\theta_{n}=\left(
N-n+1\right)  /2$ and $K_{n}=2$ for all $n$).}%
\label{fig:Nak_fig3}%
\end{center}
\end{figure}

First, we present results for the case where each hop fades according to a
Nakagami-$m$ distribution. Fig. \ref{fig:Nak_fig1} through Fig.
\ref{fig:Nak_fig3} illustrate the outage probability as a function of
$\bar{\gamma}$ for different three-hop links. Simulations are plotting along
with analytical results obtained through the Meijer $G$-function expression
(\ref{eq:Pout_Nakm_full}) as well as the leading order asymptotic result
dervied from the application of the residue theorem, which is given by
(\ref{eq:Nak_leading_order1}) and (\ref{eq:Nak_leading_order2}). Furthermore,
we compared our results to a lower bound on the probability of outage that was
developed in \cite{Karagiannidis2006a}. This bound, which can be obtained
through an application of a harmonic-geometric mean bound on the end-to-end
SNR, has subsequently been applied and discussed in a number of works
(\emph{e.g.}, \cite{Farhadi2008}).

The accuracy of our analysis is clear from these examples. Moreover, although
the asymptotic expression converges to simulation results for large
$\bar{\gamma}$, we see that the Meijer $G$-function expression is a good
approximation even at low and mid-range SNR. We also observe that the
harmonic-geometric mean bound is loose in some cases, particularly when the
inherent diversity in the channel decreases with each hop (\emph{i.e.},
$m_{n}$ decreases with increasing $n$). This behavior was also noted in
\cite{Farhadi2008}.

Fig. \ref{fig:Nak_fig3} illustrates the effect that quasi-identical fading has
on the convergence of the asymptotic expression. In this example, the first
and third hops yield the minimum shape parameter ($m_{1}=m_{3}=1$). This leads
to a second order pole in the residue analysis, and thus $P_{o,Nak}\sim
\psi_{Nak}\left(  \log\bar{\gamma}\right)  \bar{\gamma}^{-1}$. The logarithmic
term delays convergence as discussed in Section \ref{sec:convergence}, and
this is apparent in the figure in the plot of the leading order expansion. It
is also clear that the approximation can be made to be much more accurate for
moderate SNR levels by refining the expansion through the inclusion of all
correction terms of order $\Omega\left(  \bar{\gamma}^{-1}\right)  $ (the
curve labelled \textquotedblleft Refined Asymptotic\textquotedblright\ in the
figure)\footnote{The notation $f\left(  x\right)  =\Omega\left(  g\left(
x\right)  \right)  $ implies $\exists\,\,x_{0},k>0$ such that $f\left(
x\right)  \geq k\cdot g\left(  x\right)  $ for $x>x_{0}$.}.

\begin{figure}
[ptb]
\begin{center}
\includegraphics[
height=4.0179in,
width=5.1897in
]%
{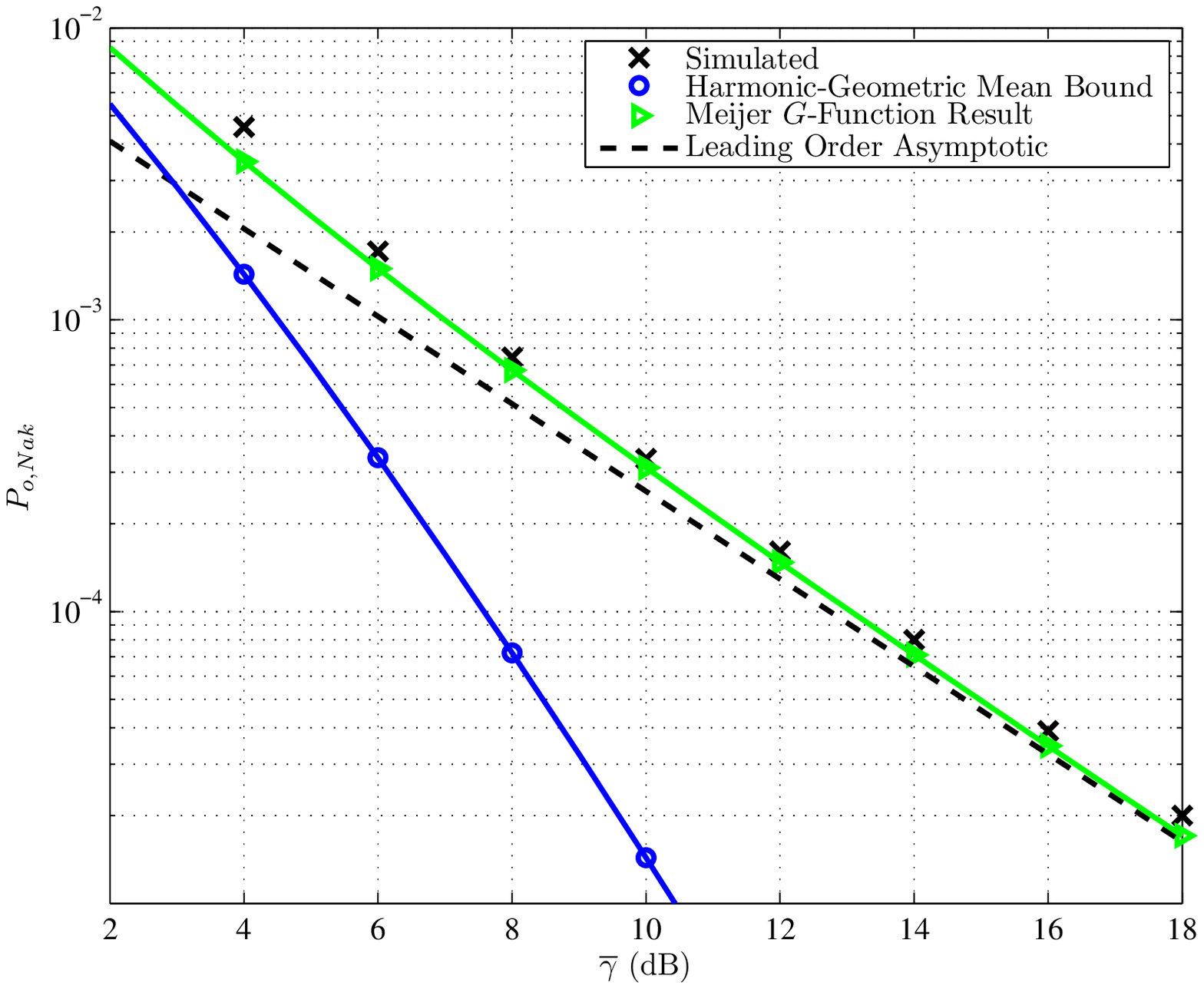}%
\caption{$P_{o,Nak}$ vs. $\bar{\gamma}$ for a fixed-gain AF multihop systems
with Nakagami-$m$ fading channels ($N=5$, $m_{1}=m_{2}=5$, $m_{3}=m_{4}=2.5$,
$m_{5}=1.5$; and $K_{n}=2$ and $\theta_{n}=\rho_{n}=1$ for all $n$).}%
\label{fig:Nak_fig4}%
\end{center}
\end{figure}

Finally, we provide results for a five-hop system in Fig. \ref{fig:Nak_fig4}.
Here, we adopt the same shape factors that were used in \cite[\emph{cf.} Fig.
4 therein]{Karagiannidis2006}; specifically, we have $m_{1}=m_{2}=5$,
$m_{3}=m_{4}=2.5$, and $m_{5}=1.5$. Again, there is excellent agreement
between the simulation results and the Meijer $G$-function expression for
$P_{o,Nak}$, and the asymptotic result converges around 13 dB. In contrast,
the harmonic-geometric mean lower bound diverges.

\begin{figure}
[ptb]
\begin{center}
\includegraphics[
height=4.0101in,
width=5.1603in
]%
{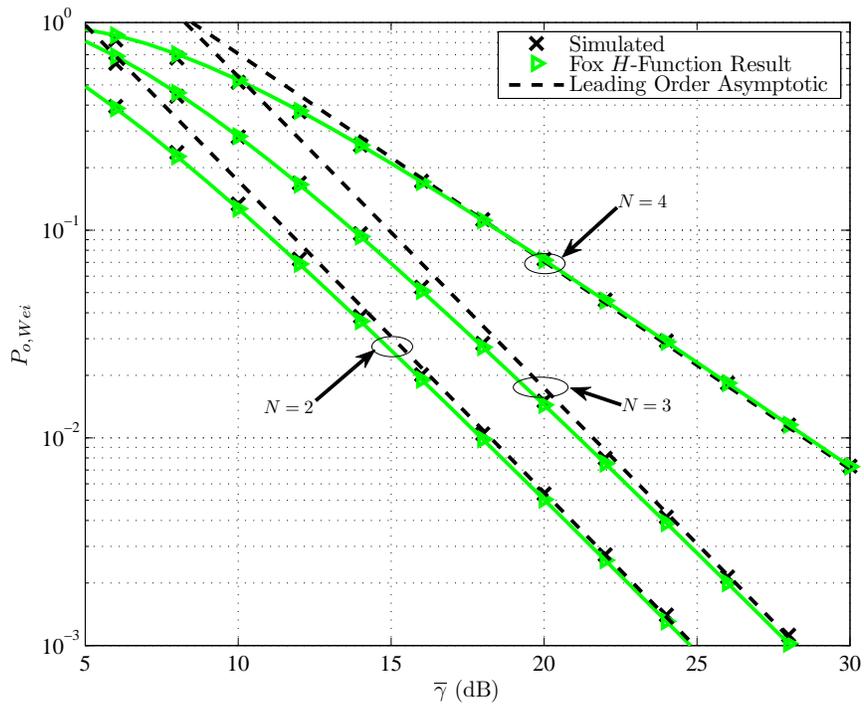}%
\caption{$P_{o,Wei}$ vs. $\bar{\gamma}$ for a fixed-gain AF multihop system
with Weibull distributed channels ($N=2,3,4$, $m_{1}=1.5$, $m_{2}=2$,
$m_{3}=2.5$, $m_{4}=1$; and $K_{n}=2$ and $\theta_{n}=\rho_{n}=1$ for all
$n$).}%
\label{fig:Wei_fig}%
\end{center}
\end{figure}

In Fig. \ref{fig:Wei_fig}, we present results for homogeneous links where each
hop experiences Weibull fading. Two, three, and four hop systems were
considered where the shape parameters were chosen to be $m_{1}=1.5$, $m_{2}=2
$, $m_{3}=2.5$, and $m_{4}=1$. Again, the analytical results agree well with
the simulations. Furthermore, we see how the addition of a fourth hop with
less inherent diversity compared to the other hops affects performance. As
previously mentioned, the hop with the minimum shape parameter dictates
performance at high SNR, which is evident from the loss in diversity in the
$N=4$ hop link in the figure.

\begin{figure}
[ptb]
\begin{center}
\includegraphics[
height=4.0101in,
width=5.1612in
]%
{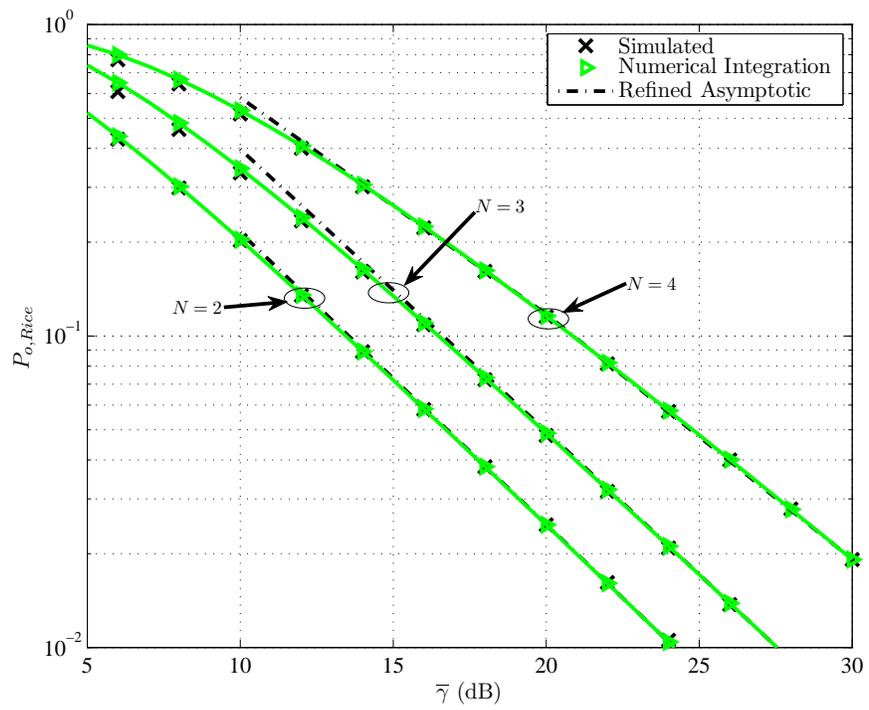}%
\caption{$P_{o,Rice}$ vs. $\bar{\gamma}$ for a fixed-gain AF multihop system
with Rician distributed channels ($N=2,3,4$, $K_{1}=1$, $K_{2}=3$, $K_{3}=5$,
$K_{4}=0$; and $K_{n}=2$ and $\theta_{n}=\rho_{n}=1$ for all $n$).}%
\label{fig:Rice_fig}%
\end{center}
\end{figure}

\begin{figure}
[ptb]
\begin{center}
\includegraphics[
height=4.0101in,
width=5.1612in
]%
{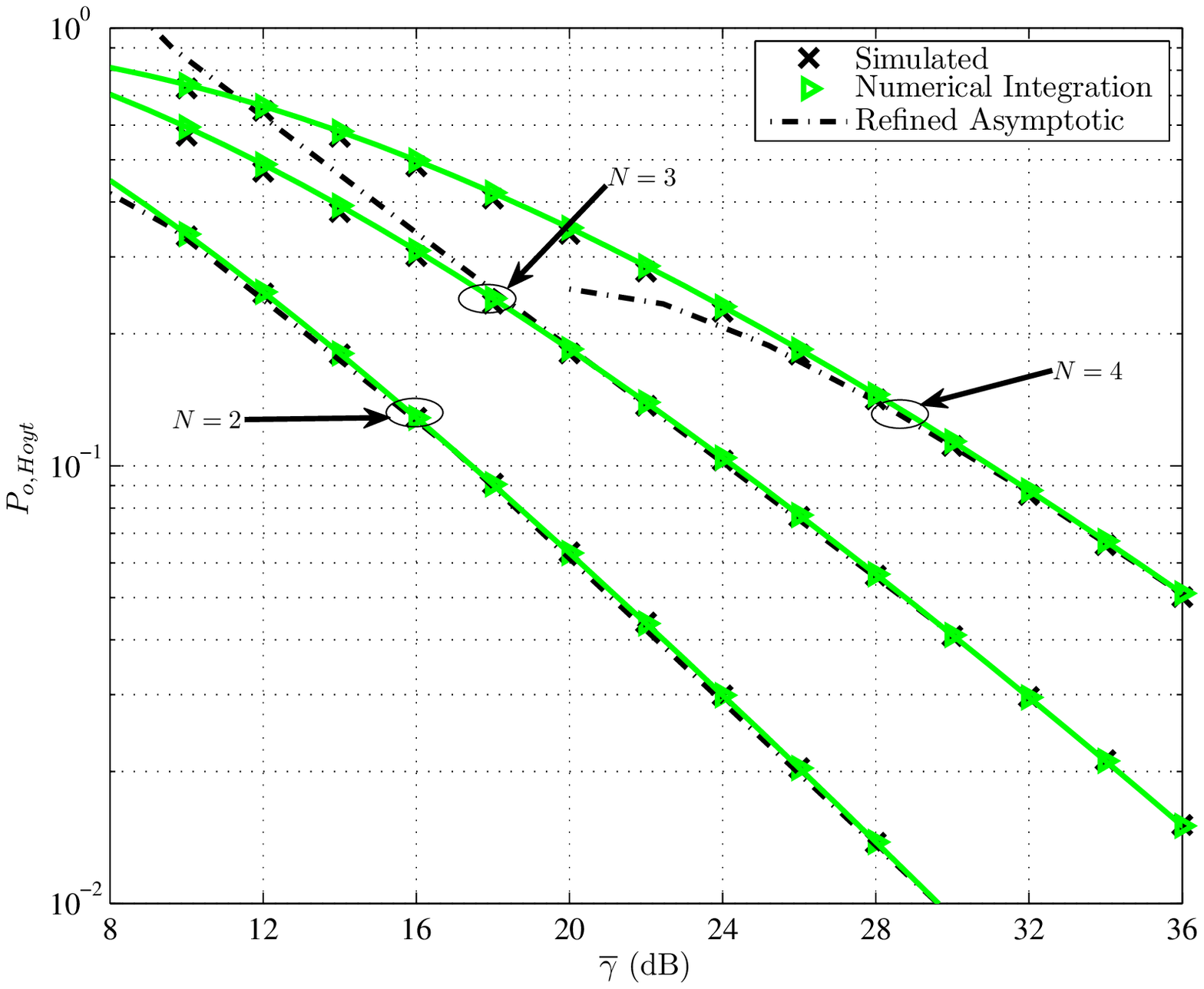}%
\caption{$P_{o,Hoyt}$ vs. $\bar{\gamma}$ for a fixed-gain AF multihop system
with Hoyt distributed channels ($N=2,3,4$, $q_{1}=3/4$, $q_{2}=1/2$,
$q_{3}=1/3$, $q_{4}=1/4$; and $K_{n}=2$ and $\theta_{n}=\rho_{n}=1$ for all
$n$).}%
\label{fig:Hoyt_fig}%
\end{center}
\end{figure}

Results for Rician and Hoyt channels are given in Fig. \ref{fig:Rice_fig} and
Fig. \ref{fig:Hoyt_fig}, respectively. The degradation in performance with
increasing numbers of hops is apparent for both cases. Moreover, the loss in
\emph{finite SNR} diversity can also be observed, which results from the
$\log\bar{\gamma}$ terms in the numerator of the high SNR expansion for
$P_{o}$. The characterization of this behavior is beyond the scope of this
paper, but is an interesting observation, nonetheless.

\begin{figure}
[ptb]
\begin{center}
\includegraphics[
height=4.0188in,
width=5.1612in
]%
{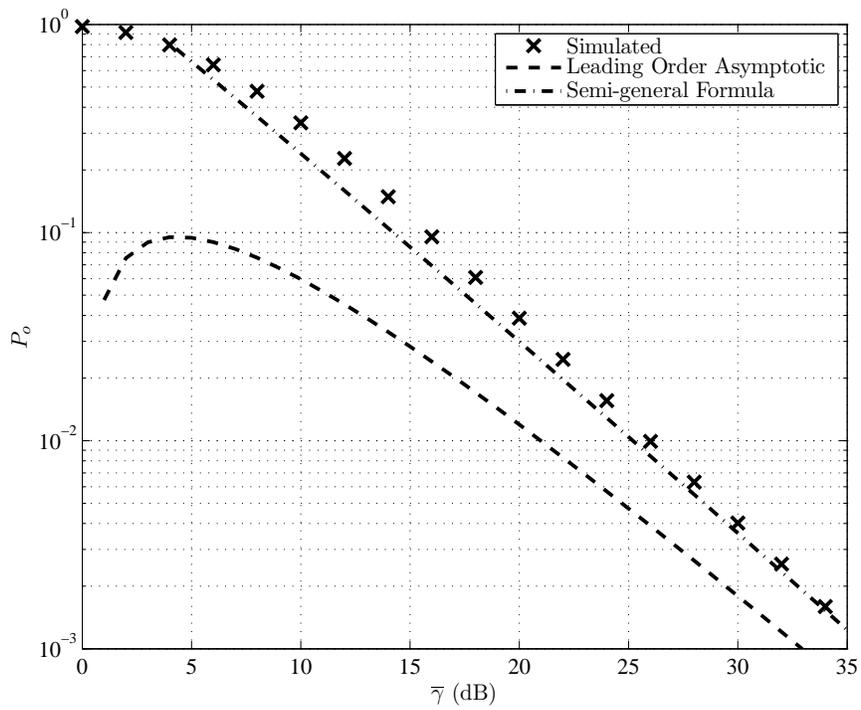}%
\caption{$P_{o}$ vs. $\bar{\gamma}$ for a four-hop inhomogeneous fixed-gain AF
multihop system. The channels corresponding to the four hops follow a
Nakagami-$m$, Weibull, Rician, and Hoyt distribution, respectively with the
following parameters: $m_{1}=2$, $m_{2}=1.5$, $K_{3}=3$, $q_{4}=3/4$,
$\theta_{n}=n/2$, and $\rho_{n}=1-\left(  n-1\right)  /10$.}%
\label{fig:inhom_fig}%
\end{center}
\end{figure}

Finally, we present results for an inhomogeneous link in Fig.
\ref{fig:inhom_fig}. In this example, the fading channel corresponding to the
first hop adheres to a Nakagami-$m$ distribution with $m_{1}=2$. The second
hop channel follows a Weibull profile with $m_{2}=1.5$. The channels related
to the third and fourth hops follow Rician and Hoyt distributions,
respectively, with $K_{3}=3$ and $q_{4}=3/4$. The scale parameter for the
$n$th hop is given by $\theta_{n}=n/2$ and we have defined $\rho_{n}=1-\left(
n-1\right)  /10$. In Fig. \ref{fig:inhom_fig}, the leading order asymptotic
and the semi-general formula are plotted along with simulation results. Again,
we see that although the graph points to convergence for the leading order
expression in the asymptotic limit, the semi-general formula provides a much
more accurate expression at low and mid-range SNR. Finally, it should be noted
that this complex example cannot be studied with the theory that has been
detailed in the literature to date, which exemplifies the versatility of our
analytical framework.

\section{Conclusions}

In this paper, we presented a novel, rigorous asymptotic analysis of the
outage probability for fixed-gain AF multihop relay systems. Our analysis was
general in nature, lending itself to application in a range of scenarios,
including cases where the per hop fading processes adhere to completely
different models and distributions. Specifically, we first provided a general
asymptotic formula for the outage probability that is applicable to any system
where the hops are statistically independent. We then provided analytical
expressions for different fading distributions -- namely Nakagami-$m$,
Weibull, Rician, and Hoyt fading -- and gave a brief discussion on the
convergence of these formulae, which culminated in a semi-general closed-form
formula for the outage probability at high SNR that can be applied to analyze
homogeneous and inhomogeneous systems. Finally, we demonstrated through
simulations that our theory is accurate, even at low to mid-range SNR in many
cases of interest.

\appendices

\section{Properties of Mellin Transforms}

The Mellin transform of a real-valued function $f\left(  x\right)  $ where
$x\geq0$ is defined by%
\[
M\left[  f;s\right]  =\int_{0}^{\infty}x^{s-1}f\left(  x\right)  \mathrm{d}x
\]
and its inverse is given by%
\begin{equation}
f\left(  x\right)  =\frac{1}{2\pi i}\int_{c-i\infty}^{c+i\infty}x^{-s}M\left[
f;s\right]  \mathrm{d}s \label{eq:invM}%
\end{equation}
for $c>0$. A useful identity exists for the Mellin transform of derivatives of
$f$, namely that \cite[6.1 (10)]{Erdelyi1954}%
\begin{equation}
M\left[  f^{\left(  \ell\right)  };s\right]  =\left(  -1\right)  ^{\ell}%
\frac{\Gamma\left(  s\right)  }{\Gamma\left(  s-\ell\right)  }M\left[
f;s-\ell\right]  \label{eq:property_1}%
\end{equation}
provided the first $\ell$ derivatives exist and are well behaved. We can also
employ Mellin transforms to approximate the average of a function (with
respect to the kernel $x^{s-1}$) evaluated at a point close to, but to the
right of $x$, \emph{i.e.}, $f\left(  x+\epsilon\right)  $ for $\epsilon>0$.
This is done by expanding $f$ near $x$, which yields%
\begin{equation}
\int_{0}^{\infty}x^{s-1}f\left(  x+\epsilon\right)  dx =\int_{0}^{\infty
}x^{s-1}\left(  \sum_{\ell=0}^{L-1}\frac{\epsilon^{\ell}}{\ell!}f^{\left(
\ell\right)  }\left(  x\right)  +\mathcal{O}\left(  \epsilon^{L}\right)
\right)  \mathrm{d}x \sim\sum_{\ell=0}^{L-1}\frac{\epsilon^{\ell}}{\ell
!}M\left[  f^{\left(  \ell\right)  };s\right]  +\mathcal{O}\left(
\epsilon^{L}\right)  \label{eq:property_2}%
\end{equation}
again, provided the derivatives exist\footnote{The notation $\sim$ denotes
asymptotic equivalence for $\epsilon \rightarrow 0$ in this case.}. Note that this is an asymptotic expansion;
convergence of the series is not guaranteed in general.

If $X$ is a random variable and $f_{X}\left(  x\right)  $
denotes its density function, then $M\left[  f_{X};s\right]  =E\left[
x^{s-1}\right]  $. \ If $f_{X}\left(  x\right)  $ is defined for $x\geq0$,
then we can define the complementary cummulative distribution function (CCDF)%
\[
\mathcal{F}_{X}\left(  x\right)  =\int_{x}^{\infty}f_{X}\left(  t\right)
\mathrm{d}t.
\]
Now we have the following identity, which is calculated using integration by
parts and, in particular, holds for probability distributions with
exponentially decaying tails\footnote{The identity is derived for the CCDF
rather than the CDF in order to ensure the transform converges.}:%
\begin{equation}
M\left[  \mathcal{F}_{X};s\right]  =s^{-1}M\left[  f_{X};s+1\right]  .
\label{eq:property_3}%
\end{equation}
Finally, suppose $Z=\prod_{n=1}^{N}X_{n}$ where $\left\{  X_{n}\right\}  $ are
statistically independent. Then it is easy to see that%
\begin{equation}
M\left[  f_{Z};s\right]  =\prod_{n=1}^{N}M\left[  f_{X_{n}};s\right]  .
\label{eq:property_4}%
\end{equation}

\section{Residue Calculations}

It is instructive to outline some of the residue calculations that were made
to obtain the results given in this paper. Most calculations follow similar
reasoning. Consequently, we only include calculations for Nakagami-$m$ fading
in this appendix.

Consider the function $I_{Nak}\left(  s\right)  $ defined in (\ref{eq:I_Nak}).
Suppose there are $T\leq N$ unique shape parameters, where the $t$th parameter
$\tilde{m}_{t}$ has multiplicity $\mu_{t}$ and $\sum_{t=1}^{T}\mu_{t}=N$. If
$\lambda_{N-1}=0$, $I_{Nak}\left(  s\right)  $ has a simple pole at $s=0$ with
residue $\operatorname{res}\left(  I_{Nak},0\right)  =\prod_{n=1}^{N}%
\Gamma\left(  m_{n}\right)  $. Moreover, $I_{Nak}$ also has a pole at
$s=s_{q,j}\triangleq m_{q}+\lambda_{q-1}+j$ for $q\in\left\{  1,\ldots
,N\right\}  $ and $j\in\left\{  0,1,\ldots\right\}  $, which is, in general,
an $r$th order pole where $1\leq r\leq N$. The residue of $I_{Nak}$ at this
pole is given by%
\begin{align}
\operatorname{res}\left(  I_{Nak},s_{q,j}\right)   &  =-\frac{1}{\left(
r-1\right)  !}\lim_{s\rightarrow s_{q,j}}\frac{\partial^{r-1}}{\partial
s^{r-1}}\left\{  \left(  s-s_{q,j}\right)  ^{r}I_{Nak}\left(  s\right)
\right\} \nonumber\\
&  =-\frac{1}{\left(  r-1\right)  !}\lim_{s\rightarrow s_{q,j}}\frac
{\partial^{r-1}}{\partial s^{r-1}}\left\{  \nu_{q,j}\left(  s\right)
z^{s}\right\} \nonumber\\
&  =-\frac{1}{\left(  r-1\right)  !}\sum_{l=0}^{r-1}\binom{r-1}{l}\nu
_{q,j}^{\left(  r-1-l\right)  }\left(  s_{q,j}\right)  \left(  \log z\right)
^{l}z^{s_{q,j}} \label{eq:rth_order_res}%
\end{align}
where the third equality follows from the Leibniz rule of differentiation of
products (with $\nu^{\left(  n\right)  }\left(  a\right)  $ being the $n$th
derivative of $\nu$ evaluated at $a$) and $\nu_{q,j}\left(  s\right)  =\left(
s-s_{q,j}\right)  ^{r}I_{Nak}\left(  s\right)  z^{-s}$.

The residue theorem states that $P_{o}$ can be expressed as a series of
residues of poles of $I_{Nak}$ \cite{Stewart1983}. Since $z\propto\bar{\gamma
}^{-1}$, if we wish to construct an approximation to $P_{o}$ that is a
function of the leading power of $z$ (as $\bar{\gamma}$ grows large), then we
can ignore all residues for which $j>0$. This leaves a finite summation of
residues corresponding to different shape parameters. In fact, it is clear
from (\ref{eq:rth_order_res}) that out of these residues, those at poles
relating to the smallest shape parameter in the set $\left\{  m_{1}%
,\ldots,m_{N}\right\}  $ dominate for large $\bar{\gamma}$. Thus, we only care
about residues of the form%
\[
\operatorname{res}\left(  I_{Nak},m_{\hat{q}}+\lambda_{\hat{q}-1}\right)
=-\frac{1}{\left(  r-1\right)  !}\sum_{l=0}^{r-1}\binom{r-1}{l}\nu_{\hat{q}%
,0}^{\left(  r-1-l\right)  }\left(  m_{\hat{q}}+\lambda_{\hat{q}-1}\right)
\left(  \log z\right)  ^{l}z^{m_{\hat{q}}+\lambda_{\hat{q}-1}}%
\]
where $\hat{q}\in\left\{  1,\ldots,N\right\}  $ such that $m_{\hat{q}}\leq
m_{n}$ for all $n$, \emph{i.e.}, $m_{\hat{q}}$ is the smallest shape
parameter, which has multiplicity $\hat{\mu}$. In particular, we must
determine conditions under which $\lambda_{\hat{q}-1}=0$ since this yields the
leading order in $z$.

Denote the ordered indices of the corresponding minimum shape parameters by
$\hat{q}_{1},\ldots,\hat{q}_{\hat{\mu}}$. Now, since $\left\{  \lambda
_{1},\ldots,\lambda_{N-1}\right\}  $ are cummulative sums of the indices
$\left\{  \ell_{1},\ldots,\ell_{N-1}\right\}  $, it follows that
$I_{Nak}\left(  s\right)  $ will have a $\hat{\mu}$th order pole at
$m_{\hat{q}}$ when $\lambda_{\hat{q}_{\hat{\mu}}-1}=0$, a $\left(  \hat{\mu
}-1\right)  $th order pole at $m_{\hat{q}}$ when $\lambda_{\hat{q}_{\hat{\mu
}-1}-1}=0$ but $\lambda_{\hat{q}_{\hat{\mu}}-1}>0$, and so on. We wish to
enumerate these instances through the indices $\left\{  \ell_{1},\ldots
,\ell_{N-1}\right\}  $. To this end, it is possible to construct $\hat{\mu}$
disjoint sets of $\left(  N-1\right)  $-tuples, which we denote $\mathcal{L}%
_{1},\ldots,\mathcal{L}_{\hat{\mu}}$, such that $\mathcal{L}_{r}$ consists of
all sets $\left\{  \ell_{1},\ldots,\ell_{N-1}\right\}  $ that satisfy the
conditions $\lambda_{\hat{q}_{r}-1}=0$ and $\lambda_{\hat{q}_{r+1}-1}>0$,
where the second condition is only necessary (and valid) for $r\leq\hat{\mu
}-1$. It follows that the order of the pole at $s=m_{\hat{q}}$ is $r$ if an
only if $\left\{  \ell_{1},\ldots,\ell_{N-1}\right\}  \in\mathcal{L}_{r}$.
This allows us to partition the expression of $P_{o}$ into terms related to
the order of the poles. Finally, we can substitute the leading order residues,
summing over the appropriate sets of $\ell$ indices, to write the following
expression for $P_{o}$%
\[
P_{o}\sim\frac{1}{\prod_{n}\Gamma\left(  m_{n}\right)  }\sum_{r=1}^{\hat{\mu}%
}\sum_{\left\{  \ell_{1},\ldots,\ell_{N-1}\right\}  \in\mathcal{L}_{r}}\left(
-1\right)  ^{\lambda_{N-1}}\xi_{\boldsymbol{\ell}}\sum_{l=0}^{r-1}\frac
{\nu^{\left(  r-1-l\right)  }\left(  m_{\hat{q}}\right)  }{\left(
r-1-l\right)  !l!}\left(  \log z\right)  ^{l}z^{m_{\hat{q}}}.
\]
Sustituting for $z$ and applying the binomial theorem yields the result given
in (\ref{eq:Pout_Nakm_expansion}).

\bibliographystyle{IEEEtran}
\bibliography{acompat,IEEEabrv,master}

\newpage%
%

\begin{table}[tbp] \centering
\caption{Mellin transforms for various fading distributions.}%
\begin{tabular}
[c]{|l|l|l|}\hline
\textbf{Distribution} & \textbf{Channel Power Density Function} $f_{X}\left(
x\right)  ,\,\,x\geq0$ & \textbf{Mellin Transform} $M\left\{  f_{X}%
;s+1\right\}  $\\\hline\hline
Nakagami-$m$ & $\theta^{-m}\Gamma\left(  m\right)  ^{-1}x^{m-1}e^{-x/\theta} $
& $\theta^{s}\Gamma\left(  s+m\right)  /\Gamma\left(  m\right)  $\\
Weibull & $m\theta^{-m}x^{m-1}e^{-\left(  x/\theta\right)  ^{m}}$ &
$\theta^{s}\Gamma\left(  s/m+1\right)  $\\
Rician & $\theta^{-1}\left(  K+1\right)  e^{-\left(  K+\theta^{-1}\left(
K+1\right)  x\right)  }I_{0}\left(  \sqrt{\frac{4K\left(  K+1\right)
x}{\theta}}\right)  $ & $e^{-K}\left(  \frac{\theta}{K+1}\right)  ^{s}%
\Gamma\left(  s+1\right)  \,_{1}F_{1}\left(  s+1,1;K\right)  $\\
Hoyt & $\frac{1+q^{2}}{2q\theta}e^{-\frac{\left(  1+q^{2}\right)  ^{2}}%
{4q^{2}\theta}x}I_{0}\left(  \frac{1-q^{4}}{4q^{2}\theta}x\right)  $ &
$\left(  \frac{2q}{1+q^{2}}\right)  ^{2s+1}\theta^{s}\Gamma\left(  s+1\right)
\,_{2}F_{1}\left(  \frac{s+1}{2},\frac{s+2}{2};1;\left(  \frac{1-q^{2}%
}{1+q^{2}}\right)  ^{2}\right)  $\\\hline
\end{tabular}
\label{tab:mellin}%
\end{table}%

\end{document}